\documentclass[a4paper]{jpconf}
\usepackage{graphicx}
\usepackage{color}

\usepackage{bm}
\usepackage{dcolumn}
\usepackage{verbatim}
\usepackage{mathbbol}
\usepackage{makeidx}
\usepackage{amsgen}
\usepackage{amsfonts}
\usepackage{amsbsy}
\usepackage{amssymb}
\usepackage{placeins}
\usepackage{cite}

\definecolor{dark_red}{rgb}{0.75,0,0}
\definecolor{dark_purple}{rgb}{0.75,0,0.75}
\definecolor{dark_blue}{rgb}{0,0,0.75}
\definecolor{dark_green}{rgb}{0,0.60,0}
\usepackage[colorlinks,linkcolor=dark_green,citecolor=blue,linkcolor=dark_green]{hyperref}

\newcommand{\XUV}{{\scriptscriptstyle\mathrm{XUV}}}

\newcommand{\IR}{{\scriptscriptstyle\mathrm{IR}}}
\newcommand{\RABITT}{{\footnotesize\textsc{RABITT }}}
\newcommand{\APT}{{\scriptscriptstyle\mathrm{APT}}}
\newcommand{\HH}{{\scriptscriptstyle\mathrm{H}}}
\newcommand{\SB}{{\scriptscriptstyle\mathrm{SB}}}
\newcommand{\eqref}[1]{(\ref{#1})}

\newcommand{\DipoleOperatorCal}{{\mathcal{O}}}
\newcommand{\DipoleOperator}{{O}}
\newcommand{\ResonantChannel}{\alpha}
\newcommand{\NonResonantChannel}{2}

\newcommand{\uam}{Departamento de Qu\'imica, M\'odulo 13, Universidad Aut\'onoma de Madrid, 28049 Madrid, Spain, EU}
\newcommand{\imdea}{Instituto Madrile\~no de Estudios Avanzados en Nanociencia (IMDEA-Nanociencia), Cantoblanco, 28049 Madrid, Spain, EU}
\newcommand{\ifimac}{Condensed Matter Physics Center (IFIMAC), Universidad Aut\'onoma de Madrid, 28049 Madrid, Spain, EU}

\begin{document}

\bibliographystyle{unsrt}

\title{Modulation of Attosecond Beating by Resonant Two-Photon Transition}

\author{\'Alvaro Jim\'enez Gal\'an}
\address{\uam}
\ead{alvaro.jimenez@uam.es}

\author{Luca Argenti}
\address{\uam}

\author{Fernando Mart{\'i}n}
\address{\uam}
\address{\imdea}
\address{\ifimac}

\begin{abstract}
We present an analytical model that characterizes two-photon transitions in the presence of autoionising states. We applied this model to interpret resonant \RABITT spectra, and show that, as a harmonic traverses a resonance, the phase of the sideband beating significantly varies with photon energy.  This phase variation is generally very different from the $\pi$ jump observed in previous works, in which the direct path contribution was negligible. We illustrate the possible phase profiles arising in resonant two-photon transitions with an intuitive geometrical representation.
\end{abstract}

\section{Introduction}

Autoionising states (AI) are hallmarks of electronic correlation, which shapes the reactivity of all many-body systems. AI states have been the subject of extensive investigation since Madden and Codling reported the asymmetric profile of helium doubly excited states in the first energy-resolved x-ray photoabsorption spectrum recorded using synchrotron radiation, a pioneering experiment which signed the birth of modern photoelectron spectroscopy [1]. Several years before, Fano had developed a model [2] in which he explained the asymmetric profiles in atomic photoelectron spectra, such as those seen by Madden and Codling, as interferences between two one-photon paths, a direct one, from the ground state to the continuum, and an indirect one, from the ground to the metastable state to the continuum. Synchrotron radiation gave access to the study of one-electron processes with unprecedented detail; the width and energy of several autoionising states have been accurately measured. Due to its typical properties (incoherent light pulses, with a duration of several picoseconds, that are highly monochromatized before impinging on the sample), however,  synchrotron radiation did not provide access to the energy dependence of resonant transition phases. Hence, the dynamical information of the resonant process, which is encoded in such phases, was lost.

The advent of attosecond light sources opened the way to monitor and control the electron motion in atoms and molecules at their intrinsic timescale [3]. In particular, such coherent ultra-fast sources of light provide a new means to investigate the role of AI states in atomic transitions since, first, the coherence of the pulses used permits to experimentally access the phases of the transitions and, second, the attosecond resolution permits to follow the temporal evolution of the AI states on a time scale smaller than their lifetime. The technique of reconstruction of attosecond beating in two-photon transitions (\RABITT) [4], which has already been successfully employed to study resonant transitions [5,6], is particularly indicated to this task. So far, however, only transitions through either electronic bound states [5] or autoionising vibronic states without any appreciable contribution from the intermediate continuum (direct path) [6] have been considered. In those conditions, the phase of the complex transition amplitude is expected to undergo a change of $\pi$ as a function of the detuning of the pump harmonic from the intermediate resonant state, and the measurements were indeed found to be compatible with this expectation.
There remained a need to explore the case in which both the bound and the continuum intermediate states contributed to the two-photon transition. To do so, it was necessary to extend Fano's model to the multi-photon finite-pulse formulation required by modern attosecond interferometric techniques. A two-photon finite-pulse resonant model, in particular, serves as a framework to interpret the phases of resonant transition amplitudes in current experiments.

In this paper, we provide a derivation of the latter theoretical model, which was originally presented in [7], and which is able to reproduce to a great accuracy time-resolved resonant two-photon transitions. Indeed, the model has already been successfully used to interpret two recent \RABITT experiments on resonant transitions in Helium [8] and Argon [9]. We also illustrate a geometrical construction that permits to interpret the phase profiles obtained from experiments. For a more complete set of results and a more detailed derivation, the reader is referred to Refs.~[7,9,10].

\section{Theory}
The lifetimes of autoionising states are often comparable to the duration of the APT and of the IR pulses used in \RABITT spectroscopy. Furthermore, the central frequency of the harmonics may not coincide with a nominal multiple of $\omega_\IR$. For these reasons, to make quantitative predictions it is necessary to use a finite-pulse formulation of the the perturbative transition amplitudes. In the present section, we provide a derivation for the time-dependent two-photon resonant model, in the simplified case of only one isolated intermediate resonance. The derivation accounts also for a direct dipolar coupling between the localised component of the intermediate metastable state and the final continumm. 

The time-dependent lowest-order perturbative amplitude for the two-photon transition from an initial atomic bound state $|g\rangle$ to a final non-resonant continuum state $|\gamma E\rangle$, for a linearly polarized field $\vec{F}(t)=\hat{\epsilon} F(t)$, is
\begin{equation}\label{eq:convolution}
\mathcal{A}_{\gamma E,g} = -\mathrm{i} \int d\omega \tilde{F}(E-E_g-\omega) \tilde{F}(\omega)\mathcal{M}_{\gamma E,g}(\omega),
\end{equation}
where $\mathcal{M}_{\gamma E,g}(\omega)=\langle \gamma E | {\DipoleOperatorCal} G_0^+(\omega_g+\omega) {\DipoleOperatorCal}| g\rangle$ is the two-photon dipole matrix element, the Fourier transform (FT) is defined as $\tilde{F}(\omega) = (2\pi)^{-1/2}\int F(t)\exp(i \omega t)dt$, $G_0^+(E)\equiv (E-H+i0^+)^{-1}$ is the retarded resolvent of the field-free hamiltonian $H$, ${\DipoleOperatorCal}=\hat{\epsilon}\cdot\vec{\DipoleOperator}$ is the dipole operator (in velocity gauge, $\vec{F}$ is the vector potential and $\vec{\DipoleOperator}=\alpha\vec{P}$,  where $\alpha$ is the fine-structure constant). The index $\gamma$ indentifies collectively all the quantum numbers, other than energy, that uniquely identify the final continuum, and which are needed to differentiate it from other continuum states, such as those populated by one-photon transition from the ground state.
Atomic units are used throughout, unless otherwise stated.
We assume that the field of the impinging pulses can be expressed as a linear combination of Gaussian pulses, 
$F(t)= F_0 \exp[-\sigma^2(t-t_0)^2/2]\cos[\omega_0 (t-t_0)+\phi]$, where $F_0$, $\omega_0$, $t_0$, $\sigma$ and $\phi$ are the amplitude, carrier frequency, center, spectral width and carrier-envelope phase of the pulse, respectively. The FT of such pulse is $\tilde{F}(\omega)=\tilde{F}^+(\omega) + \tilde{F}^-(\omega)$, where $\tilde{F}^{\pm}(\omega) = F_0 (2\sigma)^{-1}\exp(i \omega t_0)\exp[-(\omega \mp \omega_0)^2/(2\sigma^2)]\exp(\mp i\phi)$
are the components responsible for photon absorption and emission, respectively.  In a pump-probe experiment with an XUV attosecond-pulse train (APT) in association with an isolated IR pulse, the APT center $t_\XUV$ conventionally defines the time origin, $t_\XUV=0$, while the center of the IR pulse coincides with the pump-probe time delay $\tau$, $t_\IR=\tau$. In the present context, the total external field is conveniently expressed in terms of synchronized pulses for the odd harmonics, which give rise to the APT, and of a delayed pulse for the IR field, $F(t,\tau)=F_{\APT}(t)\,\,+\,\,F_{\IR}(t-\tau)$, where $F_{\APT}(t)=\sum_{n} F_{\HH_{2n+1}}(t)$.
In the \RABITT scheme, where one XUV photon is absorbed and one IR photon is either absorbed or emitted, the transition amplitude corresponding to a given sideband {\footnotesize\textsc{SB}}$_{2n}$ is given by the sum of four contributions, each associated to a time-ordered perturbative diagram. Since the contribution of the two diagrams in which the IR photon is exchanged first is generally small and can be neglected, so the transition amplitudes become $\mathcal{A}_{\gamma E,g}^{\pm,\HH_{2n\mp1}} = -\mathrm{i} \int d\omega \tilde{F}^{\pm}_\IR(E-E_g-\omega) \tilde{F}^{+}_{\HH_{2n\mp1}}(\omega)\mathcal{M}_{\gamma E,g}(\omega)$.
The energy-resolved intensity of the sideband, $I_\SB$, is computed from the transition amplitudes as 
\begin{equation}
I_{\SB}=\left|\mathcal{A}^{+,\HH_{2n-1}}_{\gamma E,g}\right|^2+\left|\mathcal{A}^{-,\HH_{2n+1}}_{\gamma E,g}\right|^2
+\,2\,\Re \left[\mathcal{A}^{+,\HH_{2n-1}*}_{\gamma E,g}\mathcal{A}^{-,\HH_{2n+1}}_{\gamma E,g}\right]
\end{equation}

To evaluate the two-photon ionization matrix element $\mathcal{M}_{\gammaÊE,g}(\omega)$, we assume that the intermediate eigenstates of $H$ are well represented by a single resonant channel, $H|\psi_{\ResonantChannel E}\rangle=|\psi_{\ResonantChannel E}\rangle E$ that can be expressed, following Fano's formalism~[2], in terms of known bound state, $|a\rangle$, and featureless continuum states, $|\ResonantChannel\varepsilon\rangle$, which are eigenstates of a reference hamiltonian $H_0$, $H_0|a\rangle=E_a|a\rangle$, $H_0|\ResonantChannel E\rangle=|\ResonantChannel E\rangle\varepsilon$, $\langle\alpha E|\alpha E'\rangle=\delta(E-E')$,
\begin{eqnarray}
|\psi_{\ResonantChannel E}\rangle &=& |\ResonantChannel E\rangle + 
\left(|a\rangle+\int \frac{d\varepsilon|\ResonantChannel\varepsilon\rangle\,V_{\ResonantChannel\varepsilon, a}}{E-\varepsilon+i0^+}
\right) \frac{V_{a,\ResonantChannel E}}{E-\tilde{E}_a},\quad\langle\psi_{\ResonantChannel E}|\psi_{\ResonantChannel E'}\rangle=\delta(E-E'),\label{eq:FanoIntermediate}
\end{eqnarray}
where $V_{a,\ResonantChannel E}=\langle a|H-H_0|\ResonantChannel E\rangle$, $\tilde{E}_a=\bar{E}_a-i\Gamma_a/2$ is the complex resonance energy, 
$\bar{E}_a=E_a+P\int d\varepsilon |V_{a,\ResonantChannel\varepsilon}|^2/(E-\varepsilon)$, $\Gamma_a=2\pi|V_{a,\ResonantChannel E}|^2$~[2].  For simplicity, in the present derivation we disregard the role played by virtual excitations of intermediate bound states.
The transition matrix elements between a localised state and Fano continuum can be parametrized as $\langle\psi_{\ResonantChannel E}|{\DipoleOperatorCal}|g\rangle={\DipoleOperatorCal}_{\ResonantChannel E, g}(\epsilon_{E}+q)(\epsilon_{E}-i)$, where $ \epsilon_{E}=2(E-\bar{E}_a)/\Gamma_a$ and $q={\DipoleOperatorCal}_{\tilde{a}g}/(\pi V_{aE}{\DipoleOperatorCal}_{Eg})$~[2].
The two-photon ionization matrix element, therefore, can be written as 
\begin{eqnarray}
\mathcal{M}_{\gamma E,g}(\omega)&=&\int  \frac{d\varepsilon\langle\gamma E | {\DipoleOperatorCal} | \psi_{\ResonantChannel \varepsilon}\rangle}{\omega_g+\omega-\varepsilon + i0^+}
\frac{\epsilon_{\varepsilon}+q}{\epsilon_{\varepsilon}-i}{\DipoleOperatorCal}_{\ResonantChannel \varepsilon, g}.\label{eq:Mc2}
\end{eqnarray}
By applying the \emph{on shell} approximation, $\langle \gamma E|{\DipoleOperatorCal}|\ResonantChannel \varepsilon\rangle\simeq\bar{\DipoleOperator}_{\gamma\ResonantChannel}(E)\delta(E-\varepsilon)$, which, for singly-charged parent ions, is quite accurate at energies of the order of 1~a.u. above the threshold or larger, and assuming that both $\bar{{\DipoleOperatorCal}}_{\gamma\ResonantChannel}(E)$ and $V_{a,\ResonantChannel E}$ are sufficiently slowly varying functions of $E$, it is immediate to see that 
\begin{equation}
\langle\gamma E| {\DipoleOperatorCal} | \psi_{\ResonantChannel \varepsilon}\rangle=\bar{{\DipoleOperatorCal}}_{\gamma\ResonantChannel}\delta(E-\varepsilon)+
\frac{\bar{{\DipoleOperatorCal}}_{\gamma\ResonantChannel}}{\varepsilon-E+i0^+}\frac{1}{\pi(\epsilon_{\varepsilon}+i)}+
\frac{{\DipoleOperatorCal}_{\gamma E,a}}{\pi V_{\ResonantChannel E,a}(\epsilon_{\varepsilon}+i)}.
\end{equation} 
We can now insert this expression in the two-photon matrix element, 
\begin{equation}
\mathcal{M}_{\gamma E,g}(\omega)=
\frac{\bar{{\DipoleOperatorCal}}_{\gamma\ResonantChannel}{\DipoleOperatorCal}_{\ResonantChannel E, g}}{\omega_g+\omega-E + i0^+}
\frac{\epsilon_{E}+q}{\epsilon_{E}-i}+
\int\frac{(\epsilon_{\varepsilon}+q)/(\epsilon_{\varepsilon}^2+1)}{\omega_g+\omega-\varepsilon+i0^+}
\left(\frac{\bar{{\DipoleOperatorCal}}_{\gamma\ResonantChannel}}{\varepsilon-E+i0^+}+\frac{{\DipoleOperatorCal}_{\gamma E,a}}{V_{\ResonantChannel E,a}}\right)
\frac{{\DipoleOperatorCal}_{\ResonantChannel \varepsilon, g}}{\pi}d\varepsilon\nonumber
\end{equation}
The integral in this last expression can be easily computed closing the integration circuit in the lower half of the complex plane and applying Cauchy's residual theorem, 
\begin{equation}
\mathcal{M}_{\gamma E,g}(\omega)=
\frac{\epsilon_{E}+q}{\epsilon_{E}+i}
\frac{\bar{{\DipoleOperatorCal}}_{\gamma\ResonantChannel}{\DipoleOperatorCal}_{\ResonantChannel E, g}}{\omega_g+\omega-E + i0^+}+
\left(\beta_a-\frac{1}{\epsilon_{E}+i}\right)
(q-i)\frac{\bar{{\DipoleOperatorCal}}_{\gamma\ResonantChannel}{\DipoleOperatorCal}_{\ResonantChannel E, g}}{\omega-\omega_{\tilde{a}g}},
\end{equation}
where $\omega_{ij}\equiv\omega_i-\omega_j$ and we introduced the parameter $\beta_a=\pi {\DipoleOperatorCal}_{\gamma E,a}V_{a,\ResonantChannel E}/\bar{{\DipoleOperatorCal}}_{\gamma\ResonantChannel}$.

If one is interested in the long-pulse limit only, for which the harmonic spectrum is strongly peaked at $\omega_\HH$, the resonant and non-resonant two-photon transition matrix elements, for the absorption of an XUV photon $\omega_\HH$ followed by the absorption/emission of one IR photon, $\omega_\IR$, can be approximated with their value at $\omega=\omega_\HH$, 
\begin{equation}
\mathcal{M}_{\gamma E,g}^{\pm}(\omega)\simeq M^{\pm}_{\gamma E,g}\,
\frac{\epsilon_\HH+q^\pm}{\epsilon_\HH+i},\qquad
q^{\pm}=q\mp(q-i)\zeta_a,\qquad\zeta_a\equiv\frac{2\beta_a\omega_\IR}{\Gamma_a}
\end{equation}
where we used the energy-conservation principle $E=E_g+\omega_\HH\pm\omega_\IR$ and we introduced the notation $\epsilon_\HH=\epsilon_{E_g+\omega_\HH}$ for the reduced detuning of the harmonic from the resonance.
Notice that when the intermediate autoionizing state $|a\rangle$ is not directly radiatively coupled to the final continuum, i.e., $\beta_a=0$, we recover an expression equivalent to Eq.~(4) in the main manuscript,
\begin{equation}
\mathcal{M}_{\gamma E,g}^{\pm}(\omega)\,\,\simeq\,\, M^{(\ResonantChannel)\pm}_{\gamma E,g}\,\,
\frac{\epsilon_\HH+q}{\epsilon_\HH+i}\,\,
+\,\,M^{(\NonResonantChannel)\pm}_{\gamma E,g}.
\end{equation}

The dimensionless parameter $\zeta_a=2\omega_\IR\beta_a/\Gamma_a={\DipoleOperatorCal}_{\gamma E,a}/(\bar{{\DipoleOperatorCal}}_{\gamma\ResonantChannel}\omega_\IR^{-1} V_{\ResonantChannel a})$ expresses the relative strength of the direct $|a\rangle\to|\gamma E\rangle$ radiative transition compared to the indirect path for the same process $|a\rangle\to|\ResonantChannel E\rangle\to|\gamma E\rangle$, in which the system first decays non-radiatively to the continuum  and subsequently exchanges a photon in a continuum-continuum transition.
\begin{figure}[hbtp!]
\centering
\includegraphics[scale=0.33]{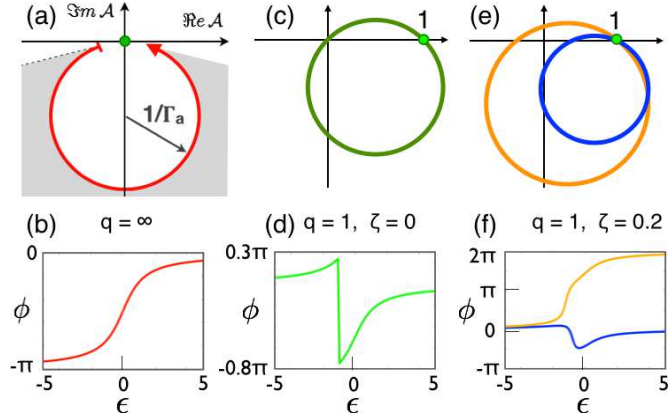}
\caption{\label{fig:SchemeCircles} Trajectory of the resonant two-photon amplitude in the complex plane (upper panels) and corresponding phase (lower panels), as the reduced detuning $\epsilon=2(E-\bar{E}_a)/\Gamma_a$ increases from large negative to large positive values. In the left panels, the only contribution to the transition comes from the intermediate bound component ($q=\infty$). In the central panels, both the intermediate continuum and bound components contribute to the dipolar transition from the ground state ($q=1$), but the intermediate bound component is not radiatively coupled to the final continuum $\zeta_a=0$. Finally, in the right panels, this final restriction is removed ($\zeta_a=0.2$), and, as a consequence, the circle of the transition amplitude to the upper sideband is contracted and misses the origin, thus experiencing no net phase change, while the circle of the transition amplitude to the lower sideband gets expanded, thus encircling the origin and experiencing a net phase transition as large as $2\pi$. See text for more details.}
\end{figure}
As $\epsilon_\HH$ increases from $-\infty$ to $+\infty$ (i.e., from large negative to large positive detuning of the resonant harmonic) the resonant factor $(\epsilon_\HH+q^\pm)/(\epsilon_\HH+i)$ describes, counterclockwise, a circular trajectory in the complex plane, which starts from $(1,0)$, is centered at $(1-iq^{\pm})/2$ and has radius $|1-iq^{\pm}|/2$,
\begin{equation}
\frac{\epsilon_\HH+q^\pm}{\epsilon_\HH+i}=\frac{1}{2}(1-iq^\pm)+\frac{1}{2}(1+iq^\pm)e^{2i\phi_\HH},\quadÊ\phi_\HH\equiv\arctan(\epsilon_\HH)+\pi/2.
\end{equation}
The circle intercepts the origin only if $\zeta_a=0$. If $\zeta_a\gtrless 0$, the origin falls outside the circle and hence the phase of the resonant matrix element $\mathcal{M}_{\gamma E,g}^{\pm}$ experiences a continuous excursion with no net variation, while if $\zeta_a\lessgtr 0$, the circle encloses the origin, and the phase of $\mathcal{M}_{\gamma E,g}^{\pm}$ undergoes a smooth overall excursion of $2\pi$. In the long-pulse limit, therefore, when $\zeta_a\neq 0$, if the phase of the resonant amplitude for the absorption of one IR photon performs a jump of $2\pi$, that for the emission of an IR photon will experience no net variation, and viceversa.
 
 With finite pulses,  the transition amplitude is given by the convolution \eqref{eq:convolution} of $\mathcal{M}_{\gamma E,g}(\omega)$ with the FTs of the field. 
The calculation for Gaussian pulses is lengthy but straightforward, and the result can be expressed in closed form in terms of the Faddeeva special function, $w(z)$, here defined as the analytic continuation of $i\pi^{-1}\int_{-\infty}^{\infty}dt\exp(-t^2)/(z-t)$ ($\Im z>0$).
In the case of the absorption of an XUV photon $\omega_\HH$ followed by the absorption ($+$) or emission ($-$) of an IR photon $\omega_\IR$, the result is
\begin{equation}
\mathcal{A}_{\gamma E,g}^{\pm,\HH (\ResonantChannel)}=\mathcal{F}(\tau)\,e^{\pm i\omega_\IR\tau}\bar{\DipoleOperator}_{\gamma\ResonantChannel}(E){\DipoleOperatorCal}_{\ResonantChannel E, g}\,\left[
\frac{\epsilon_{E}+q}{\epsilon_{E}+i}w(z_E^\pm)+\left(\beta_a-\frac{1}{\epsilon_{E}+i}\right)(q-i) w(z_{\tilde{E}_a}^\pm)\right],\nonumber
\end{equation}
where $\mathcal{F}(\tau)$ is an inessential form factor of the field, and 
\begin{equation}
z_\varepsilon^{\pm}\equiv \frac{\sigma_t}{\sqrt{2}}[E-(\varepsilon\pm\omega_\IR)]-\frac{[E-(E_g+\omega_\HH\pm\omega_\IR)]/\sigma_\HH^2+i\tau}{\sqrt{2}\sigma_t},\qquad\sigma_t=\sqrt{\sigma_\HH^{-2}+\sigma_\IR^{-2}}.
\end{equation}
\begin{figure}[hbtp!]
\centering
\includegraphics[scale=0.33]{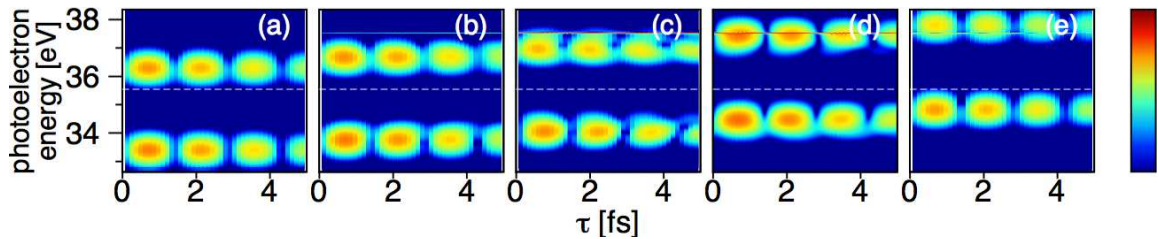}
\caption{\label{fig:Sidebands}  Sideband photoelectron signals in a \RABITT pump-probe photoionization of the helium atom from the ground state, as a function of both the photoelectron energy (vertical axis) and the pump-probe time delay (horizontal axis). Each panel corresponds to a different reduced detuning $\epsilon_\HH$ of harmonic H41 from the $2s2p$ intermediate {$^1$P$^o$} doubly excited state ($\bar{E}=35.56$~eV, $\Gamma=0.037$~eV, $q=-2.77$, $\zeta/\omega_{\IR}=0.19$~eV$^{-1}$), whose energy is indicated by the thin white dashed line: (a) $\epsilon_\HH=-43$, (b) $\epsilon_\HH=-25$, (c) $\epsilon_\HH=0.9$, (d) $\epsilon_\HH=17$, (e) $\epsilon_\HH=35$. At large negative (a) or positive (f) detunings, the upper and lower sidebands are in phase to a very good approximation. As the harmonic traverses the resonance, on the other hand, the beating of the upper and lower sidebands get clearly out of phase and the sideband profile itself is distorted. These panels are computed with an extension of the model described in the text, which accounts for final resonant states. The thin horizontal feature above 37~eV, visible in panels (c-e) is due to the final $2p^2$ {$^1$S$^e$} autoionising state.}
\end{figure}

It is instructive to verify that, in the limit of long ($\sigma_t\Gamma_{a}\gg1$) overlapping ($\tau\ll\sigma_t$) pulses, we recover expressions similar to the ones seen above, as expected. Indeed, in this limit, the argument of the Faddeeva function diverges, so one can use the asymptotic expansion $w(z) \simeq i\pi^{-1/2}z^{-1}$. At the nominal centre of the two-photon signal ($E=E_g+\omega_\HH\pm\omega_\IR$), therefore, $z^{\pm}_\varepsilon\simeq\sigma_t\left(E_g+\omega_\HH-\varepsilon\right)/\sqrt{2}$. Using the energy-preserving condition, one can simplify the transition amplitudes to
\begin{equation}
\mathcal{A}_{\gamma E, g}^{\pm,\HH}=\frac{\sqrt{2}\,i}{\pi\sigma_t} \mathcal{F}(\tau)\,e^{\pm i\omega_\IR\tau}
\mathcal{M}_{\gamma E,g}^{\pm}(\omega).
\end{equation}
With finite pulses, the resonant $\mathcal{A}^{\pm,\HH(\ResonantChannel)}_{\gamma E,g}$ amplitudes describe contracted circles. For sufficiently small values of $|\zeta_a|$, therefore, both amplitudes would fail to enclose the origin, and hence the jump of $2\pi$ wouldn't be observable.

\section{Conclusions}
We have derived a model which permits us to predict the phase variation of two-photon resonant amplitudes, extracted from current attosecond pump-probe experiments. In the limit of long pulses, we find an expression similar in form to that for the resonant one-photon transitions, but which requires an additional parameter $\zeta$ arising from the exchange of an IR photon with the autoionising state. The model has a compact analytical finite-pulse formulation, which has been instrumental to interpret recent findings in resonant \RABITT experiments [8,9].
With the use of the model and simple geometrical constructions, we have characterised the possible profiles of the amplitude phase that can be expected in resonant two-photon transitions as a function of detuning. These can vary from a finite excursion with no net phase variation to a total jump of 2$\pi$, depending on the relative strength and phases of the atomic transitions involved, the duration of the pulses used and the number of intermediate open channels.

\ack
We thank A. L'Huillier, A. Maquet, R. Ta\"ieb, P. Sali\`eres, E. Lindroth, M. Dahlstr\"om, M. Kotur, V. Gruson, T. Carette, D. Kroon, L. Barreau, J. Caillat, M. Gisselbrecht, and C. L. Arnold for useful discussions.
We acknowledge computer time from the CCC-UAM and Marenostrum Supercomputer Centers and financial support from the European Research Council under the European Union's Seventh Framework Programme (FP7/2007-2013)/ERC grant agreement 290853 XCHEM, the MINECO project FIS2013-42002-R, the ERA-Chemistry Project PIM2010EEC-00751, the European COST Action XLIC CM1204, the Marie Curie ITN CORINF, and the CAM project NANOFRONTMAG.

\section*{References}

$[1]$ R. Madden and K. Codling. Phys. Rev. Lett. {\bf 10}, 516 (1963).\\
$[2]$ U. Fano, Phys. Rev. {\bf 124}, 1866 (1961).\\
%$[3]$ C. Ott, A. Kaldun, L. Argenti,	P. Raith,	K. Meyer,	M. Laux,	Y. Zhang,	A. Bl\"attermann,	S. Hagstotz,	T. Ding,	R. Heck,	J. Madro\~nero,	F. Mart\'in and T. Pfeifer, Nature {\bf 516}, 374 (2014).\\
%$[4]$ F. Calegari, D. Ayuso, A. Trabattoni, L. Belshaw, S. De Camillis, S. Anumula, F. Frassetto, L. Poletto, A. Palacios, P. Decleva, J. B. Greenwood, F. Mart\'in and M. Nisoli, Science {\bf 346}, 336 (2014).\\
$[3]$ F. Krausz and M. Ivanov, Rev. Mod. Phys.  {\bf 81}, 163 (2009).\\
$[4]$ P. M. Paul, E. S. Toma, P. Breger, G. Mullot, F. Aug\'e, Ph. Balcou, H. G. Muller and P. Agostini, Science {\bf 292}, 1689 (2001).\\
$[5]$ M. Swoboda, T. Fordell, K. Kl\"under, J. M. Dahlstr\"om, M. Miranda, C. Buth, K. J. Schafer, J. Mauritsson, A. L'Huillier and M. Gisselbrecht, Phys. Rev. Lett. {\bf 104}, 103003 (2010).\\
$[6]$ J. Caillat, A. Maquet, S. Haessler, B. Fabre, T. Ruchon, P. Sali\`eres, Y. Mairesse, and R. Ta\"ieb, Phys. Rev. Lett. {\bf 106}, 093002 (2011).\\
$[7]$ \'A. Jim\'enez-Gal\'an, L. Argenti, and F. Mart\'in, Phys. Rev. Lett. {\bf 113}, 263001 (2014).\\
$[8]$ V. Gruson, L. Barreau, A. Jim\'enez-Gal\'an, F. Risoud, J. Caillat, A. Maquet, B. Carr\'e, F. Lepetit, J.-F. Hergott, T. Ruchon, L. Argenti, R. Ta\"ieb, F. Mart\'in and P. Sali\`eres, \emph{manuscript in preparation} (2015).\\
$[9]$ M. Kotur, D. Guenot, A. Jim\'enez-Gal\'an, D. Kroon, E. W. Larsen, M. Louisy, S. Bengtsson, M. Miranda, J. Mauritsson, C. L. Arnold, S. E. Canton, M. Gisselbrecht, T. Carette, J. M. Dahlstr\"om, E. Lindroth, A. Maquet, L. Argenti, F. Mart\'in and A. L'Huillier, arXiv:1505.02024 (2014).\\
$[10]$ \'A. Jim\'enez-Gal\'an, F. Mart\'in, and L. Argenti, \emph{manuscript in preparation} (2015).\\

\bibliography{biblio}

\end{document}